\documentclass[12pt]{iopart}
\usepackage{graphicx}

\begin{document}

\title{Critical behavior of the SIS epidemic model with time-dependent infection rate}

\author{Nuno Crokidakis $^{1,2}$ and Marcio Argollo de Menezes $^{1,2}$}

\address
{
$^{1}$Instituto de F\'{\i}sica - Universidade Federal Fluminense \\
Av. Litor\^anea s/n \\
24210-340 \hspace{5mm} Niter\'oi - Rio de Janeiro \hspace{5mm} Brazil \\
$^{2}$National Institute of Science and Technology for Complex Systems, Brazil}

\ead{nuno@if.uff.br, marcio@if.uff.br}

\begin{abstract}
\noindent
In this work we study a modified Susceptible-Infected-Susceptible
(SIS) model in which the infection rate $\lambda$ decays exponentially
with the number of reinfections $n$, saturating after $n=l$. We find a
critical decaying rate $\epsilon_{c}(l)$ above which a finite fraction of the
population becomes permanently infected. From the mean-field solution
and computer simulations on hypercubic lattices we find evidences that the upper critical dimension is $6$ like in the SIR model, which can be mapped in ordinary percolation. 
\end{abstract}

\noindent{\it Keywords\/}: Stochastic Processes, Epidemic Models, Population Dynamics,
Monte Carlo Simulation


\maketitle

\section{Introduction}

\quad Epidemic models have long been studied in the physics community
both to infer patterns and to develop policies to stop and prevent
epidemics on natural environments \cite{perelson_review, perelson_viral_dynamics,vicsek_experimental, anderson} and as a playground
to test theoretical ideas \cite{dickman, hinrichsen,newman_SIR, bailey}. 
In the simplest models, individuals can be either
susceptible or infected and, in the latter case, can infect others
with a given infection rate. If, after some time, the infected
individual is susceptible to reinfection the model is known
susceptible-infected-susceptible (SIS). If, otherwise, infected
individuals become immune it is called susceptible-infected-recovered
(SIR) model.  Despite the simplicity of these models, they have 
succesfully modelled a varied class of diseases, like dengue
\cite{derouich,esteva,nuraini}, HIV \cite{rita,baryarama} and
influenza A (H1N1) \cite{perelson_review,perelson_viral_dynamics, gordo, katriel}.

A fundamental question in the analysis of epidemic models is whether a
disease will spread through a finite fraction of the population or
will it only affect a small fraction of the individuals. Also of great
importance is the development of efficient immunization strategies,
both answers depending fundamentally on the underlying topology of
interconnections between individuals of the population \cite{cohen}. 

Although one could create a variety of infection dynamics and network
structures, an exact mapping of epidemic models onto
reaction-diffusion models \cite{hinrichsen} and subsequent development
of operator algebras and field-theories \cite{doi,janssen,
  dp_field_theory} show that a small number of universality classes
might exist, linking different models to the same quantitative
behavior around the onset of epidemic state \cite{janssen,grassberger}.
 Scaling analysis also reveal that an upper critical
dimension might exist for each process, above which dimensional
effects are irrelevant and critical exponents and amplitude ratios are
the same as in the mean-field approximation (which mimics
infinite-dimensional systems, as the number of individuals $N\to
\infty$).

Effects of network topology might also affect the properties of
epidemic models: while on networks with randomly connected agents
there is a finite threshold separating epidemic from non-epidemic
states, networks with large connectivity fluctuations, such as
scale-free networks, lack epidemic thresholds on both SIS and SIR
models
\cite{newman_SIR,may,moreno,pastor_satorras_pre,pastor_satorras_prl,
havlin_prl,boguna}.

In this work we study a modified version of the SIS model. In
particular, we consider a decaying infection rate $\lambda$, i.e.,
each individual $j$ that recovers from the infected state in a certain
time step $t$ decreases his probability to becomes infected again in
the next time step $t+1$ in the form
$\lambda_{j}(t+1)=\epsilon\,\lambda_{j}(t)$, where $\epsilon$ is a
parameter of the model. In addition, the decrease of the infection
rate occurs only a limited number $l$ of times. This type of infectivity
has been known to occur both in plants \cite{sir_plants} and in the Simian 
Immunodeficiency Virus infection \cite{perelson_siv}. We study the model on
regular lattices and in the mean-field regime, and considering the
above-mentioned modification, our results suggest that the system
undergoes a phase transition at critical values $\epsilon_{c}(l)$
separating a phase where the disease reaches a finite fraction of the
population ($\epsilon>\epsilon_{c}$) from a phase where the disease
does not spread out ($\epsilon\leq\epsilon_{c}$). We also find that
the upper critical dimension is $6$, as in the SIR model \cite{hinrichsen,newman_SIR}.


\section{Model and Mean-Field Approach}

\quad The population is classified into two classes: Susceptible (S)
and Infected (I). The transitions between the states $S$ and $I$ occur 
according to the following automata rules:
\begin{itemize}
\item If individual $j$ is Infected at time $t$, it is Susceptible at
  time $t+1$ with probability $\alpha$;
\item If, otherwise, individual $j$ is Susceptible at time $t$, then
  with probability $\lambda$ it becomes Infected at time $t+1$ if it
  is in contact with an Infected individual at the same time;
\item Each individual $j$ starts with $\lambda(j,t=0)=\lambda_{0}$,
  but $\lambda$ depends on how many times the individual was infected
  before, i.e., if an individual $j$ performed the transition $I\to S$
  in a certain time step $t$, in the following time step $t+1$ the
  infection rate $\lambda$ will be updated in the following form
\begin{equation}\nonumber
\lambda(j,t+1)=\epsilon\,\lambda(j,t) ~,
\end{equation}
\noindent
where $\epsilon$ is a parameter less than $1$ that controls the decaying of the infection rate.
\item In addition, this decay in the infection rate occurs only a
  maximum number $l$ of times, i.e., each individual has a limited
  capacity to decrease his probability of reinfection.
\end{itemize}

This dynamics can be relevant to describe diseases such as flu, for
which our chance to be reinfected decreases with age. The above rules
define a mean-field-like system because all individuals interact with
all others. Thus, an analytic approach can be developed, based on the
standard ordinary differential equations for SIS models. We can define
$S_{k}$ as the density of Suceptible individuals that have recovered
$k$ times from the disease, with $\lambda_{k}=\epsilon^{k}\lambda_{0}$
being the corresponding infection rate of these
individuals. Considering that after some time steps (nothing to do
with the steady state of the system) all individuals will be in two
states, $I$ or $S_{l}$, we can see that the only important equation in
the evolution of the system is the equation for $S_{l}$, i.e.,

\begin{figure}[t]
\begin{center}
\vspace{0.5cm}
\includegraphics[width=0.45\textwidth,angle=0]{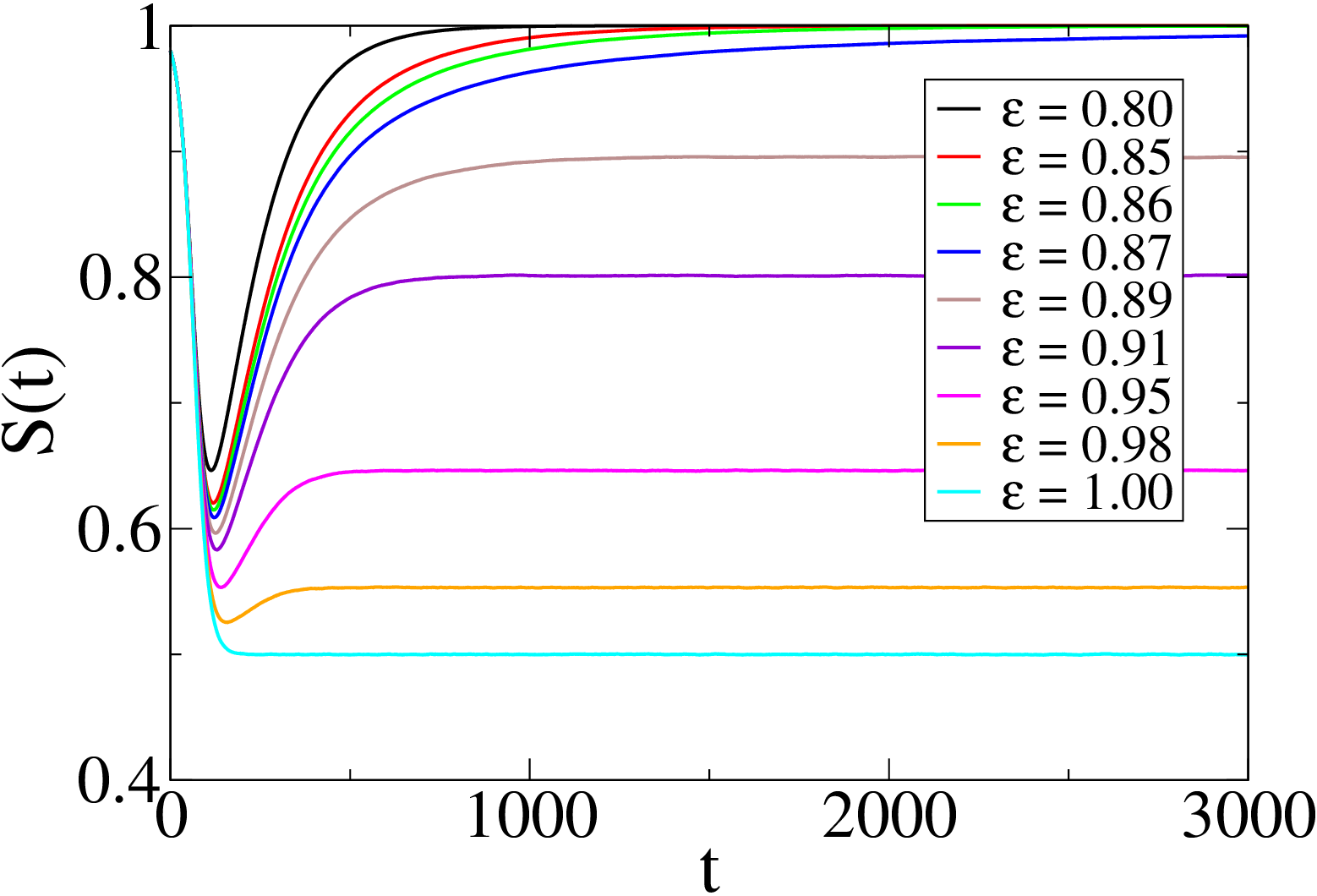}
\hspace{1.0cm}
\includegraphics[width=0.45\textwidth,angle=0]{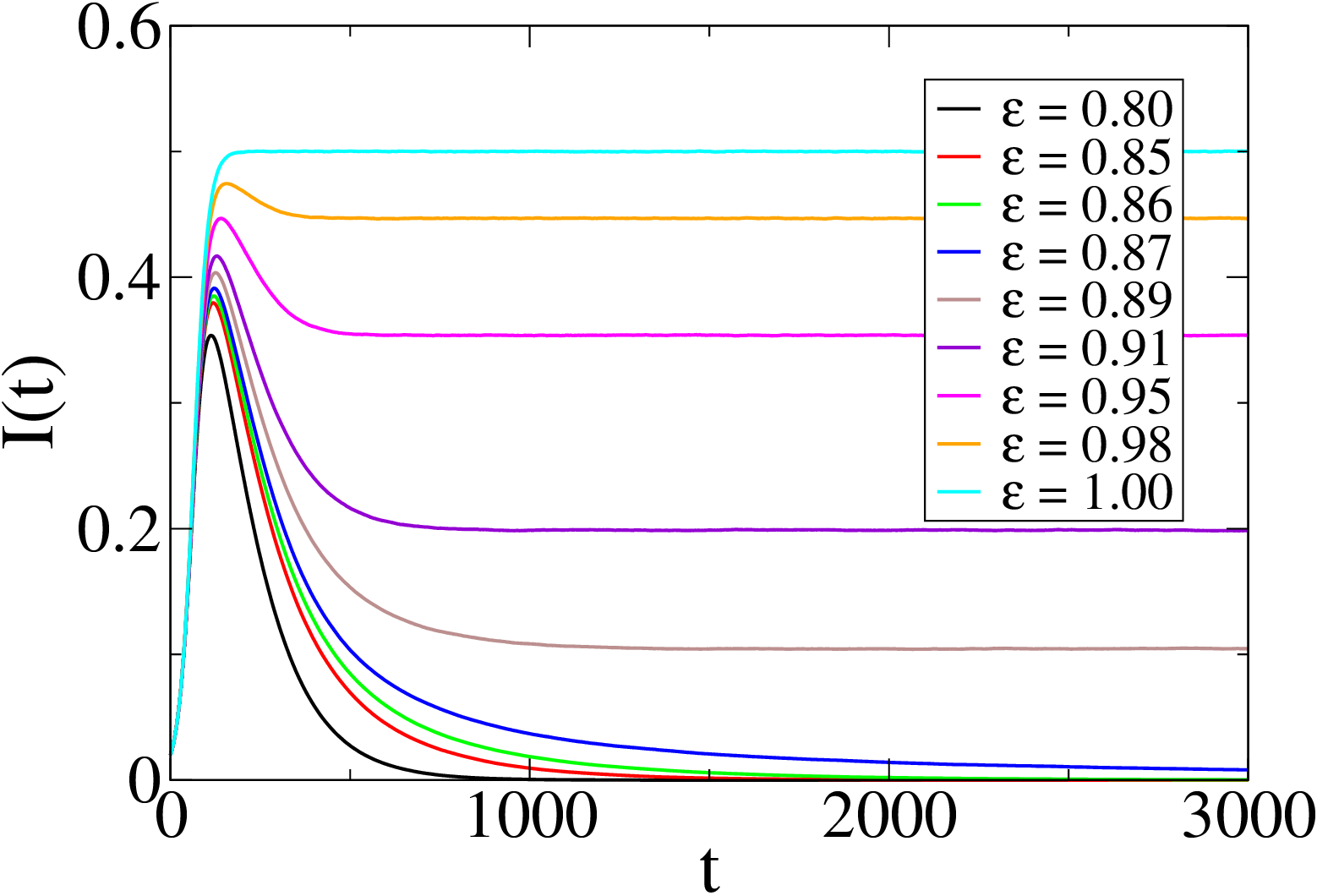}
\end{center}
\protect\caption{Density of Susceptible individuals $S(t)$ (left side,
  from top to bottom: $\epsilon=0.80, 0.85, ..., 1.00$) and density of
  Infected individuals $I(t)$ (right side, from top to
  bottom: $\epsilon=1.00, 0.98, ..., 0.80$) as functions of time $t$
  for a population of size $N=10^{5}$, limiting parameter $l=5$ and
  typical values of $\epsilon$. Data are averaged over $200$
  realizations. In this case, for $\epsilon < \sim 0.87$ the disease does
  not spread in the system. The parameters are: $\alpha=0.05$ and
  $\lambda_{0}=0.1$.}
\label{fig1}
\end{figure}

\begin{equation}\label{eq1}
\frac{dS_{l}}{dt} = \alpha I - \lambda_{l}S_{l}I ~.
\end{equation}
\noindent
Taking into account that in the steady state, i.e., for $t\to\infty$,
we have that $S_{l}=S_{{\rm steady}}$ and $dS_{l}/dt=dS_{{\rm
    steady}}/dt=0$, Eq. (\ref{eq1}) give us
\begin{equation}\label{eq2}
I_{{\rm steady}}\left(\alpha - \epsilon^{l}\lambda_{0}S_{{\rm steady}}\right)=0 ~,
\end{equation}
\noindent
where we have used the notations $S_{{\rm steady}}$ and $I_{{\rm
    steady}}$ to represent the stationary density of Susceptible and
Infected individuals, respectively, i.e., $S_{{\rm
    steady}}=S(t\to\infty)$ and $I_{{\rm
    steady}}=I(t\to\infty)$. There are two solutions of
Eq. (\ref{eq2}): $I_{{\rm steady}}=0$ and $I_{{\rm
    steady}}=1-(\alpha/\lambda_{0})\epsilon^{-l}$, where we have used
the relation $S+I=1$. Notice that the nontrivial solution for $I_{{\rm
    steady}}$ may vanishes for critical values of $\epsilon$ given by
\begin{equation}\label{eq5}
\epsilon_{c}=\left(\frac{\alpha}{\lambda_{0}}\right)^{1/l} ~.
\end{equation}
\noindent
Using this result, we can rewrite the expression for $I_{{\rm steady}}$ as
\begin{equation}\label{eq_I}
I_{{\rm steady}}=1-\left(\frac{\epsilon_{c}}{\epsilon}\right)^{l} ~.
\end{equation}

\begin{figure}[t]
\begin{center}
\vspace{0.5cm}
\includegraphics[width=0.45\textwidth,angle=0]{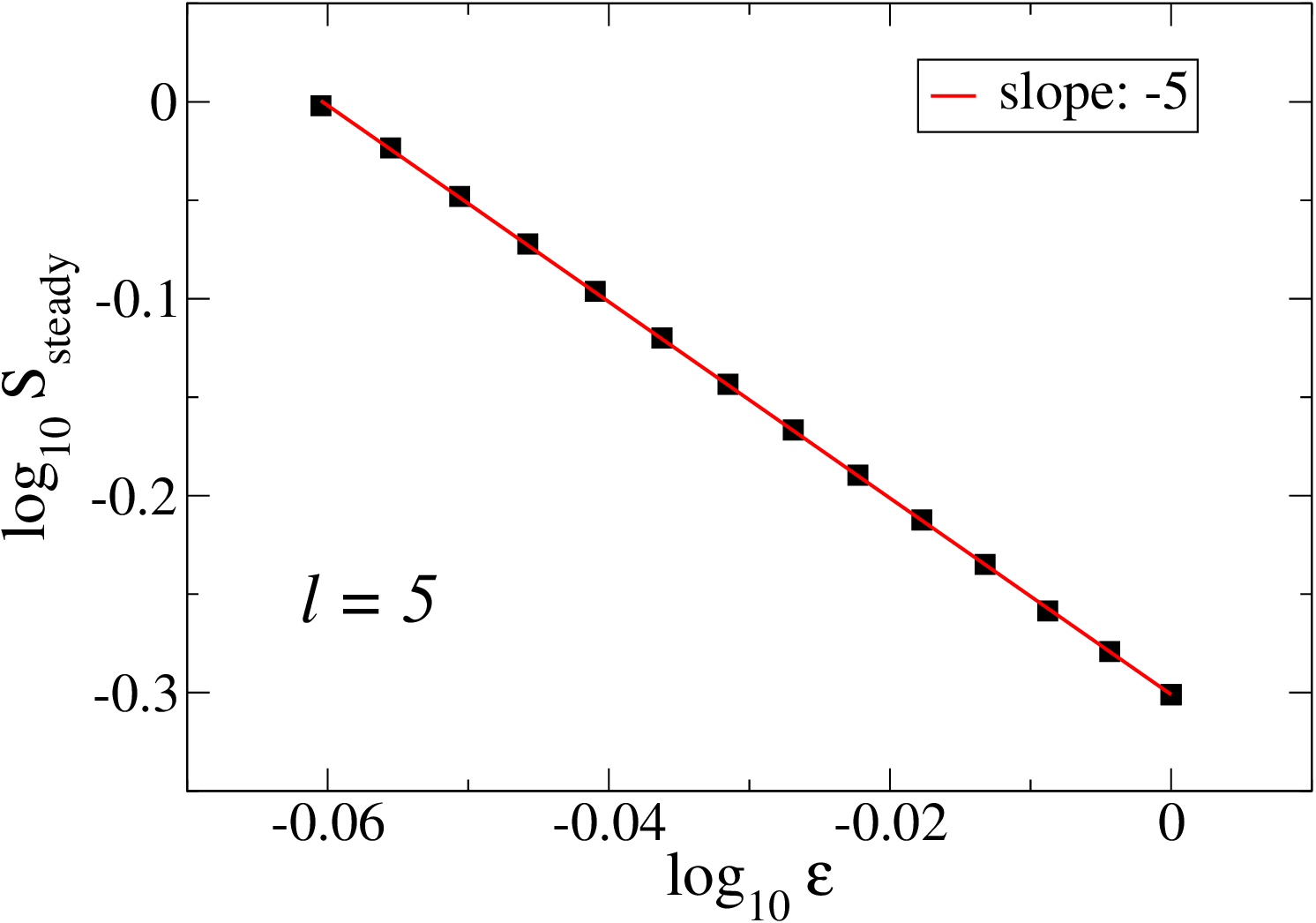}
\hspace{1.0cm}
\includegraphics[width=0.45\textwidth,angle=0]{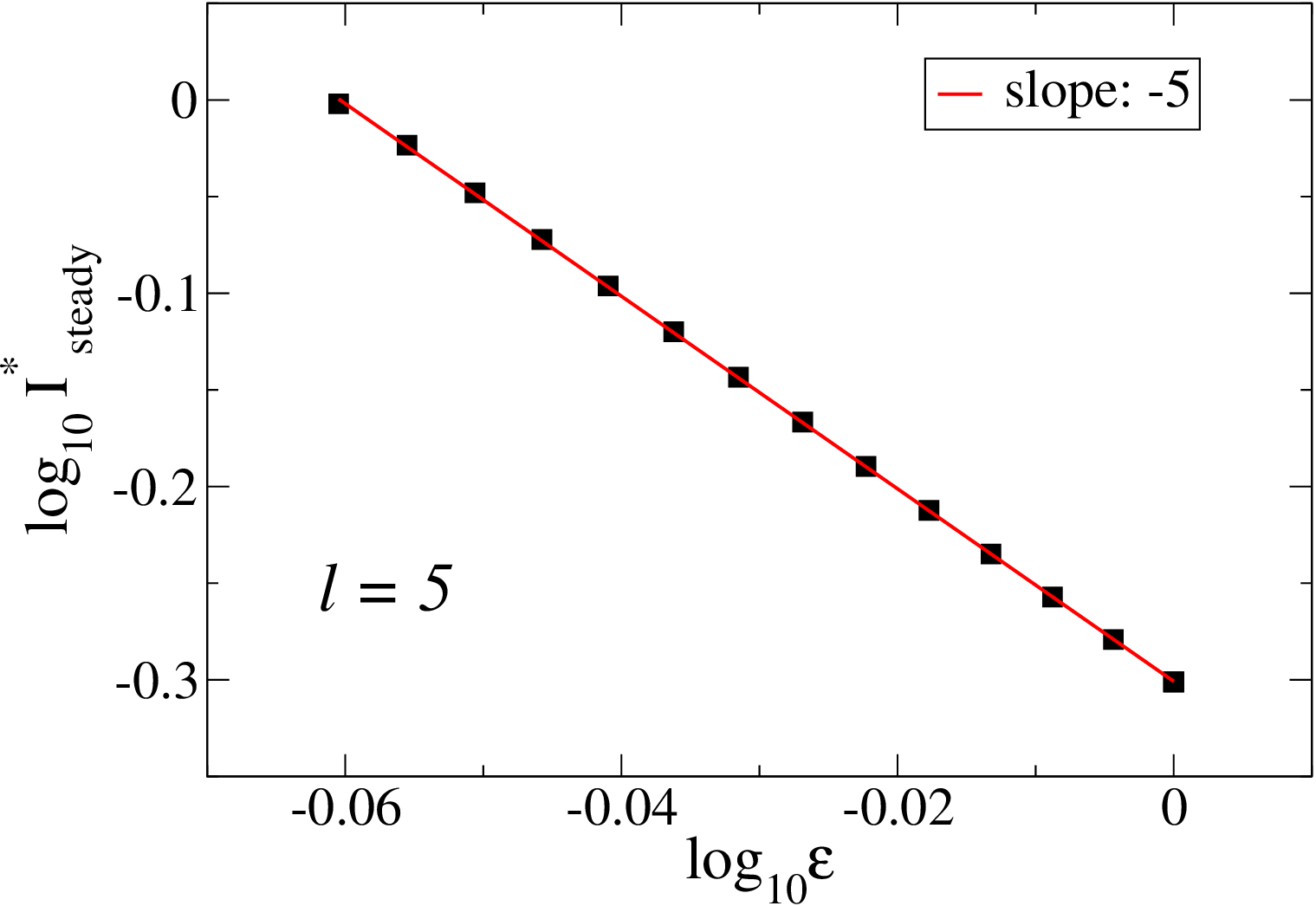}
\end{center}
\protect\caption{Stationary density of Susceptible individuals
  $S_{{\rm steady}}$ for $l=5$ and some values of $\epsilon$ in the
  log-log scale. The straight line has slope $-5$ (left side). For the
  stationary density of Infected individuals $I_{{\rm steady}}$ we do
  not have a power-law dependency on the parameter $\epsilon$, as
  predicted analytically in Eq. (\ref{eq_I}). Thus we plot here
  $I^{*}_{{\rm steady}}=1-I_{{\rm steady}}$, and the slope of the
  straight line is $-5$ (right side). The parameters are: $N=10^{5}$,
  $\alpha=0.05$ and $\lambda_{0}=0.1$. Each point is averaged over 200
  realizations.}
\label{fig2}
\end{figure}

\noindent
In other words, we have a \textit{Disease-free phase}, where the
disease disappears of the system, for $\epsilon\leq \epsilon_{c}(l)$, with
the critical values $\epsilon_{c}(l)$ given by Eq. (\ref{eq5}). On the
other hand, for $\epsilon>\epsilon_{c}(l)$ we have an \textit{Epidemic
  phase}, where the disease survives and reaches a finite fraction of
the population. Considering again the relation $S+I=1$, we can obtain
a power-law relation between the stationary density of Susceptible
individuals and the parameters of the system,
\begin{equation}\label{eq6}
S_{{\rm steady}}=\left(\frac{\alpha}{\lambda_{0}}\right)\epsilon^{-l} ~.
\end{equation}

\begin{figure}[t]
\begin{center}
\vspace{0.5cm}
\includegraphics[width=0.45\textwidth,angle=0]{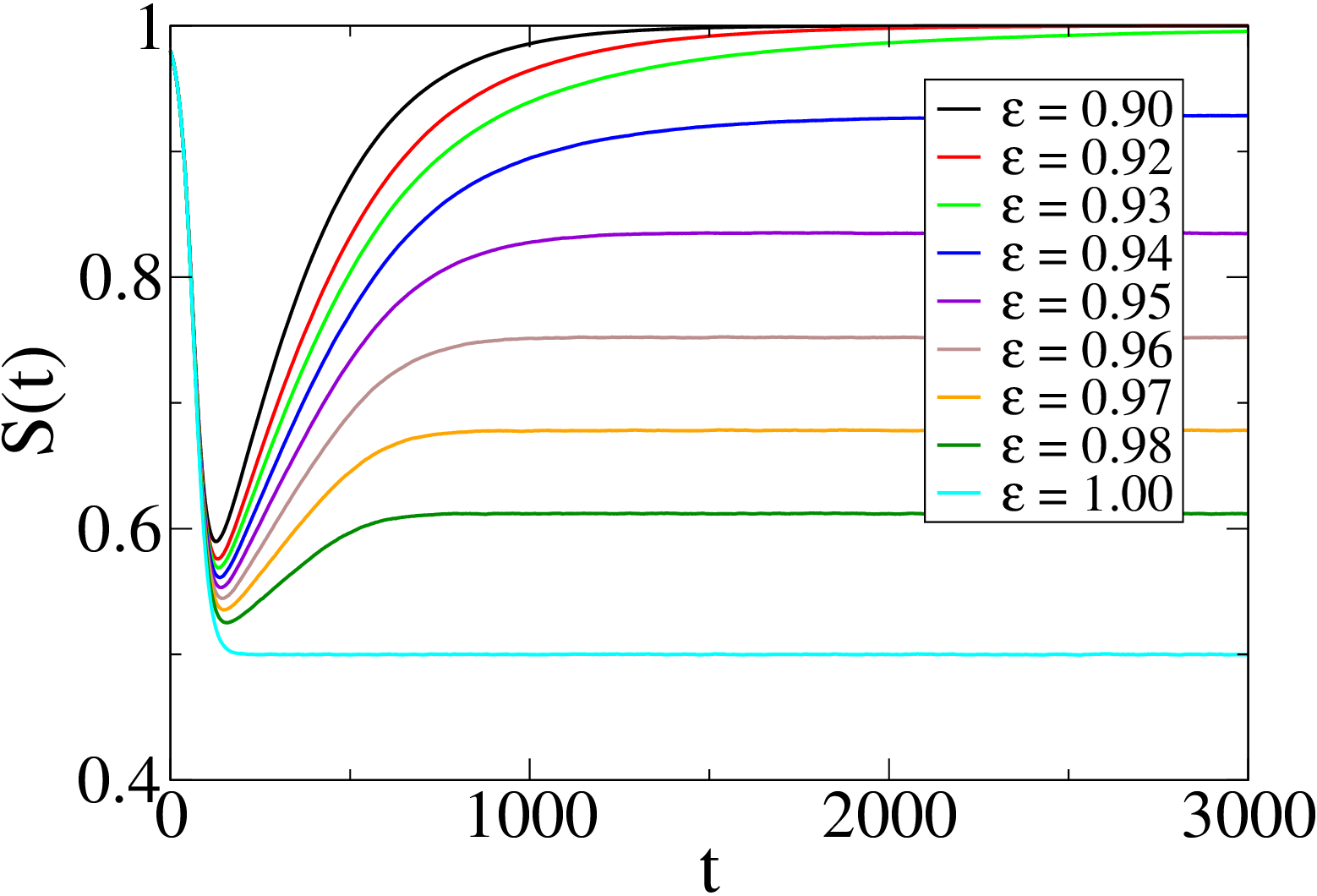}
\hspace{1.0cm}
\includegraphics[width=0.45\textwidth,angle=0]{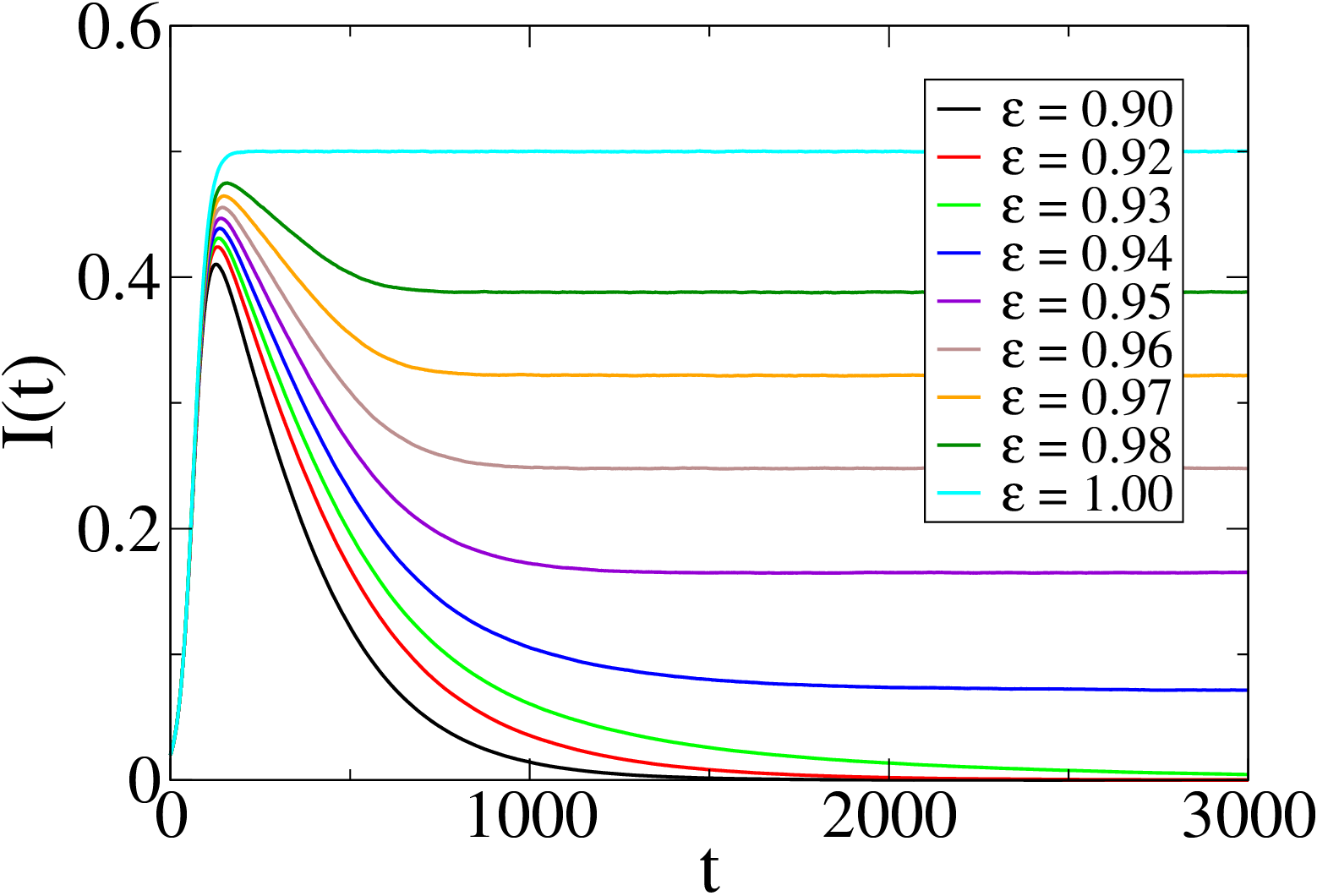}
\end{center}
\protect\caption{Density of Susceptible individuals $S(t)$ (left side,
  from top to bottom: $\epsilon=0.90, 0.92, ..., 1.00$) and density of
  Infected individuals $I(t)$ (right side, from top to bottom:
  $\epsilon=1.00, 0.98, ..., 0.90$) as functions of time $t$ for
  $N=10^{5}$ and limiting parameter $l=10$. Data are averaged over
  $200$ realizations. Notice that in this case the disease disappears
  of the system for $\epsilon < \sim 0.93$. The parameters are:
  $\alpha=0.05$ and $\lambda_{0}=0.1$.}
\label{fig3}
\end{figure}

After the development of the analytical solution of the problem, we
can confront it with Monte Carlo simulations. We simulated populations of size
$N=10^{5}$, with probabilities $\alpha=0.05$ and $\lambda_{0}=0.1$ and
different values of $\epsilon$ and $l$. We considered that $2\%$ of
the individuals are initially Infected in the population, and all results
were averaged over $200$ realizations. Following the rules presented
in the beggining of this section, the algorithm to simulate the
problem is as follows: (i) at each time step, each Infected individual
$j$ returns to the Susceptible state with probability $\alpha$; (ii)
at the same time, each Susceptible individual $j$ becomes Infected
with probability $\lambda_{j}$ if a randomly choosen node is
Infected. After each transition $I\to S$, the infection rate decreases
in the form $\lambda_{j}\to\epsilon\;\lambda_{j}$.

In Fig. \ref{fig1} we exhibit results for the density of Susceptible
individuals $S(t)$ and the density of Infected individuals $I(t)$ as
functions of the simulation time $t$ for $l=5$ and typical values of
$\epsilon$. We can observe that the system reaches steady states for
all values of $\epsilon$, and that for $\epsilon < \sim 0.87$ the
disease disappears of the system, i.e., we have $I=0$ for large
$t$. Considering the mean-field calculations derived in the beggining
of this section, the analytical prediction of Eq. (\ref{eq5}) for this
critical value is $\epsilon_{c}(l=5)\cong 0.87055$, in excellent
agreement with the numerical result.

We can analyze how the stationary values $S_{{\rm
    steady}}=S(t\to\infty)$ depend on the parameter $\epsilon$. For
this purpose, we have considered time averages of the density of
susceptible individuals (after the system reach the steady states),
and in addition these values were averaged over $200$ realizations of
the system, for each value of $\epsilon$. We can observe the power-law
behavior (see Fig. \ref{fig2}, left side)
\begin{equation}\label{eq7}
S_{{\rm steady}}\sim\epsilon^{-\nu} ~,
\end{equation}

\begin{figure}[t]
\begin{center}
\vspace{0.7cm}
\includegraphics[width=0.55\textwidth,angle=0]{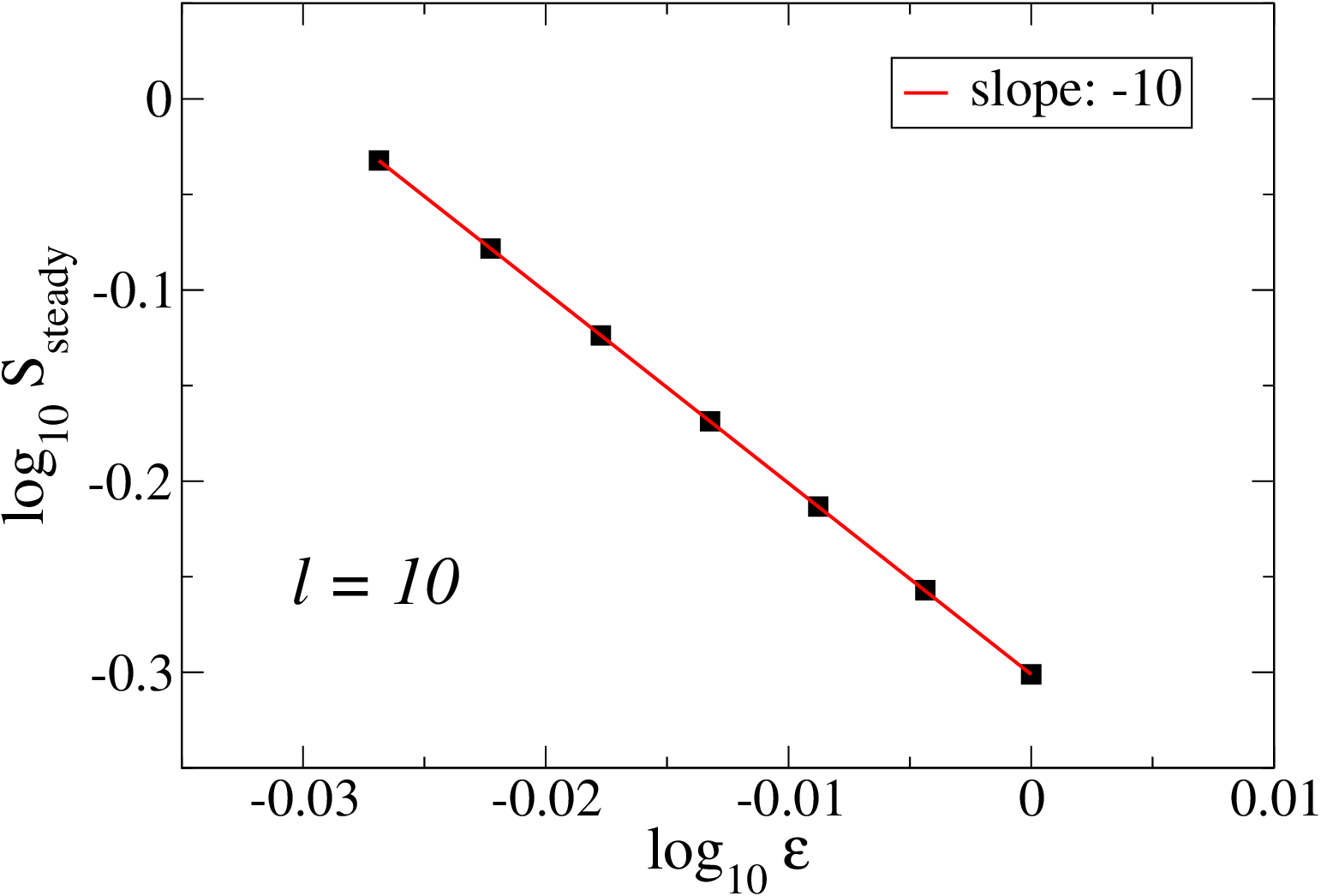}
\end{center}
\protect\caption{Stationary density of Susceptible individuals
  $S_{{\rm steady}}$ for $l=10$ and some values of $\epsilon$ in
  the log-log scale. The straight line has slope $-10$. The parameters
  are $\alpha=0.05$ and $\lambda_{0}=0.1$. Each point is averaged over
  200 realizations.}
\label{fig4}
\end{figure}

\noindent
with $\nu=5$ in this case ($l=5$). Notice that this result is in
agreement with the analytical prediction, Eq. (\ref{eq6}). On the
other hand, for the stationary values $I_{{\rm steady}}=I(t\to\infty)$
we do not have a power-law behavior, as predicted analytically in
Eq. (\ref{eq_I}). However, we can analyze the behavior of $I^{*}_{{\rm
    steady}}=1-I_{{\rm steady}}$, for which Eq. (\ref{eq_I}) give us
\begin{equation}\label{eq_II}
I^{*}_{{\rm steady}}=\left(\frac{\epsilon_{c}}{\epsilon}\right)^{l} ~,
\end{equation}
or in other words, $I^{*}_{{\rm steady}}\sim\epsilon^{-l}$, the same
behavior observed for $S_{{\rm steady}}$ [see Eq. (\ref{eq6})]. Thus,
in Fig. \ref{fig2} (right side), we exhibit the simulation data for
$I^{*}_{{\rm steady}}$ versus $\epsilon$ for $l=5$. Fitting data, we
obtained $I^{*}_{{\rm steady}}\sim\epsilon^{-5}$ \footnote{Since the
  behavior of $S_{{\rm steady}}$ and $I^{*}_{{\rm steady}}$ as
  functions of $\epsilon$ is the same, in the following we will
  analyze only the stationary density of susceptible individuals,
  $S_{{\rm steady}}$.}, in agreement with the analytical result,
Eq. (\ref{eq_II}).

In Fig. \ref{fig3} we show results for $l=10$ and different values of
$\epsilon$. Again, the system reaches steady states for all
$\epsilon$, but the disease spreads in the system only for $\epsilon > \sim
0.93$. Observe that this critical value $\epsilon_{c}(l=10)$ is
greater than the value $\epsilon_{c}(l=5)\sim 0.87$, as expected, due to
the greater capacity of the individuals to decrease their reinfection
rates. The analytical result of Eq. (\ref{eq5}) for this critical
value is $\epsilon_{c}(l=10)\cong 0.93303$, again in excellent
agreement with the numerical result.

\begin{figure}[t]
\begin{center}
\vspace{0.5cm}
\includegraphics[width=0.45\textwidth,angle=0]{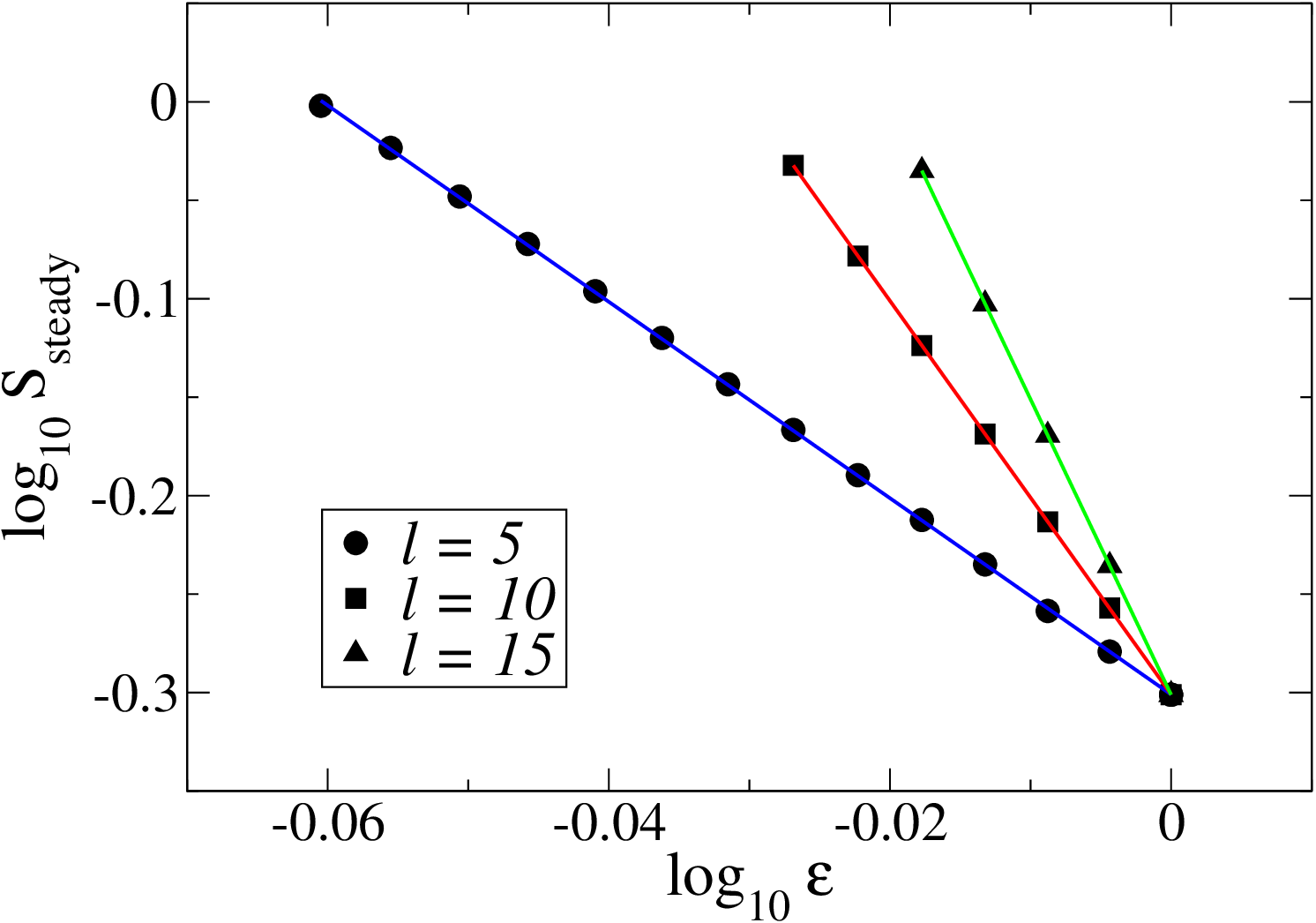}
\hspace{1.0cm}
\includegraphics[width=0.45\textwidth,angle=0]{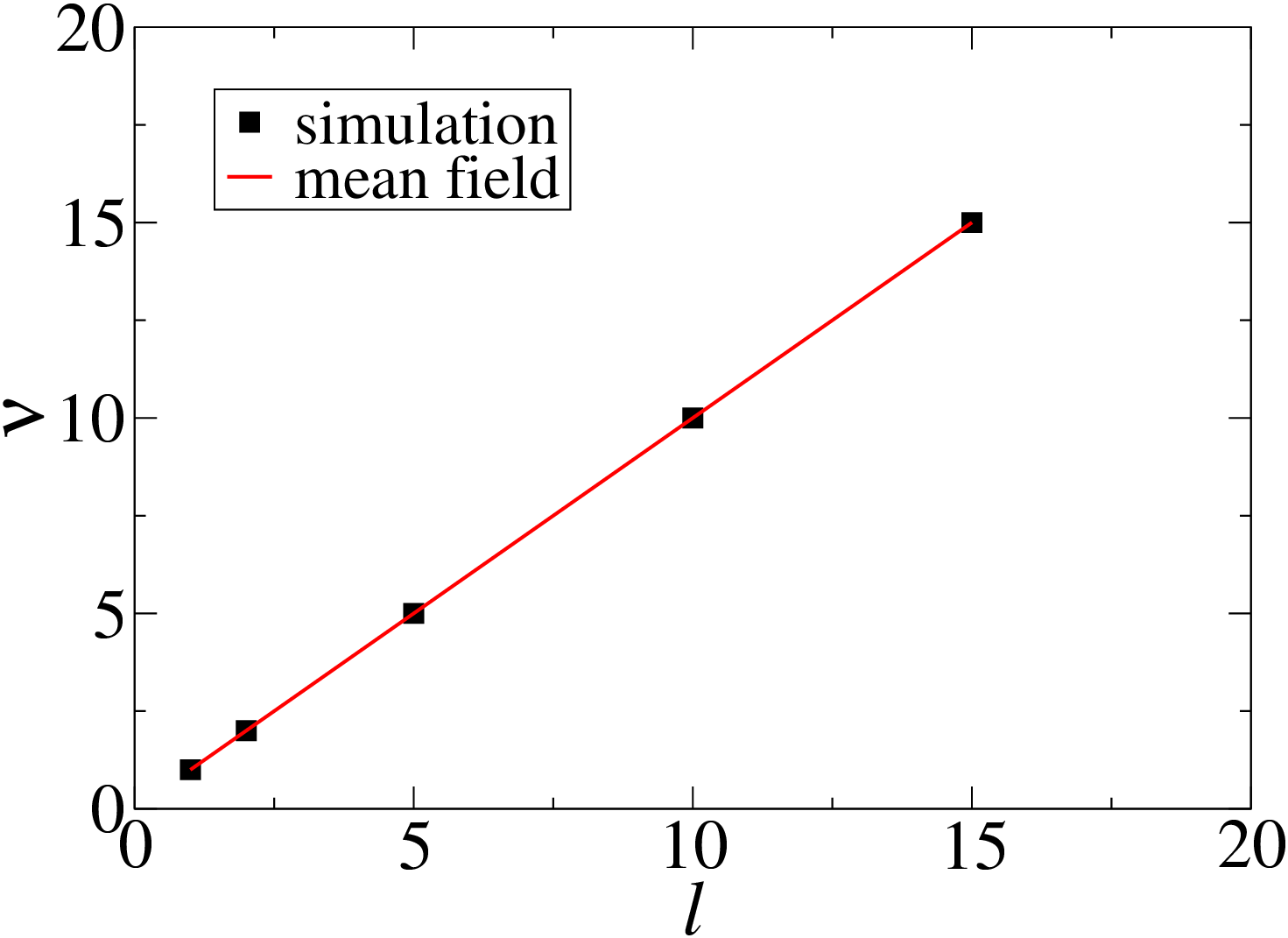}
\end{center}
\protect\caption{Stationary density of Susceptible individuals
  $S_{{\rm steady}}$ as a function of $\epsilon$ for different values
  of $l$. The straight lines are fittings, which give us $S_{{\rm
      steady}}\sim \epsilon^{-\nu(l)}$, with $\nu(l)=l$ (left
  side). It is also shown the exponent $\nu$ as a function of $l$
  (right side). The squares were estimated from the fittings, whereas
  the line is the analytical prediction, Eq. (\ref{eq6}). The
  parameters in both figures are $\alpha=0.05$ and $\lambda_{0}=0.1$.}
\label{fig5}
\end{figure}

Considering the stationary values $S_{{\rm steady}}$ for different
values of $\epsilon$, we can also observe a power-law behavior
$S_{{\rm steady}}\sim\epsilon^{-\nu}$ (see Fig. \ref{fig4}), but now
with a different exponent, $\nu=10$. In other words, we have the
general form
\begin{equation}\label{eq8}
S_{{\rm steady}}\sim\epsilon^{-\nu(l)} ~,
\end{equation}
\noindent
with $\nu(l)=l$, which is supported by numerical results for other
values of $l$ (see Fig. \ref{fig5}, left side). In addition, the
analytical prediction of Eq. (\ref{eq6}) give us the same behavior of
the above numerical result, Eq. (\ref{eq8}), as we can see in
Fig. \ref{fig5} (right side).

\begin{figure}[t]
\begin{center}
\vspace{0.5cm}
\includegraphics[width=0.55\textwidth,angle=0]{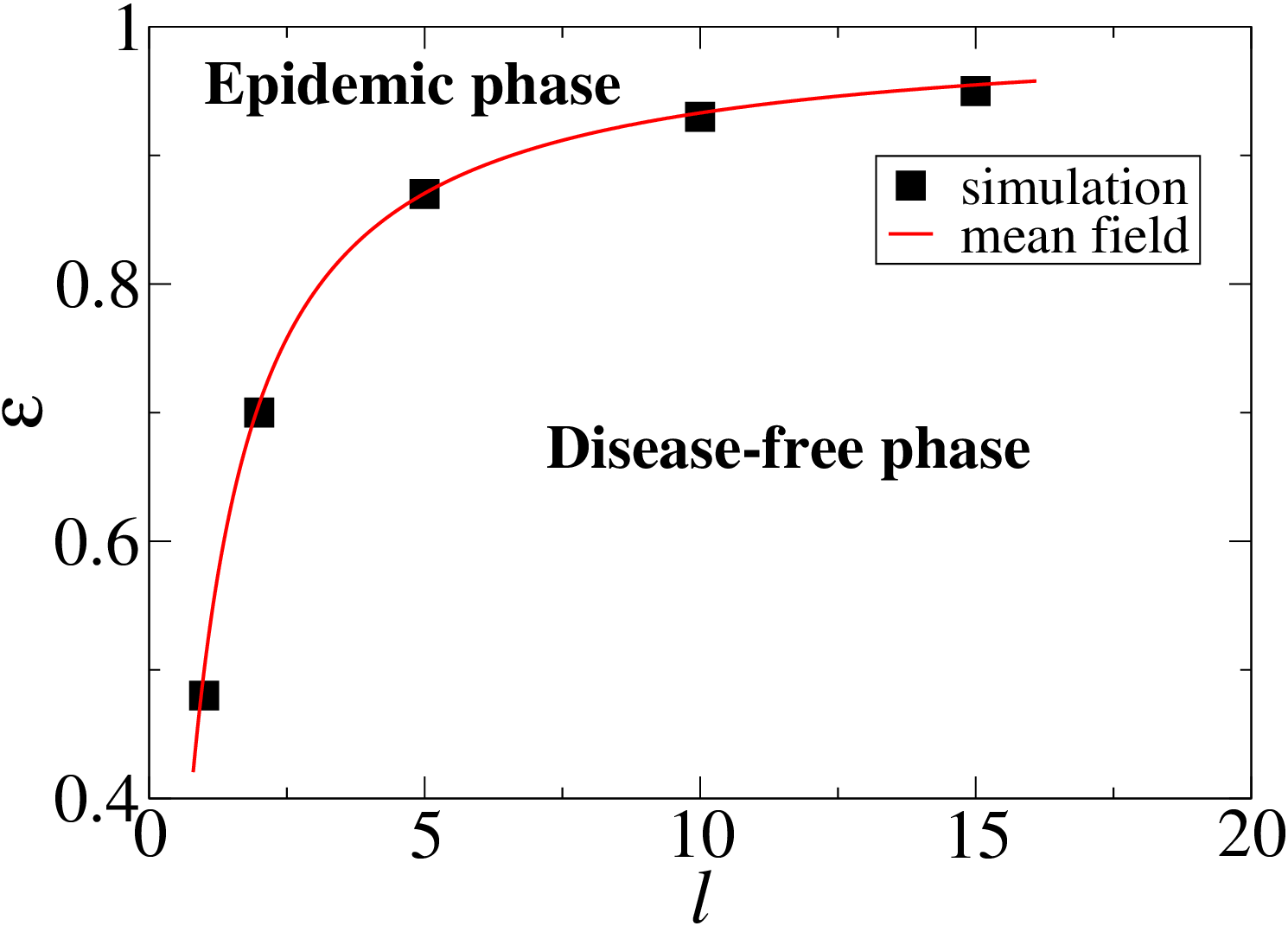}
\end{center}
\protect\caption{Phase diagram of the model in the plane $\epsilon$
  versus $l$, separating the Disease-free and the Epidemic phases. The
  squares were estimated from the simulation data, whereas the full
  (red) line is the mean-field prediction, Eq. (\ref{eq5}). The
  parameters are $\alpha=0.05$ and $\lambda_{0}=0.1$.}
\label{fig6}
\end{figure}

Taking into account the numerical results for $\alpha=0.05$,
$\lambda_{0}=0.1$ and different values of the parameters $\epsilon$
and $l$, we show in Fig. \ref{fig6} the phase diagram of the model
separating the Disease-free and the Epidemic phases. The squares are
the critical values $\epsilon_{c}(l)$ estimated from the simulation
data, whereas the curve is the analytical prediction of
Eq. (\ref{eq5}). We can observe an excellent agreement between the
analytical and the Monte Carlo results.

In the next section we will analyze how the presence of a regular topology
affects the system and the mean-field results presented in this section.

\section{Simulations on regular d-dimensional lattices}

\begin{figure}[t]
\begin{center}
\vspace{0.5cm}
\includegraphics[width=0.45\textwidth,angle=0]{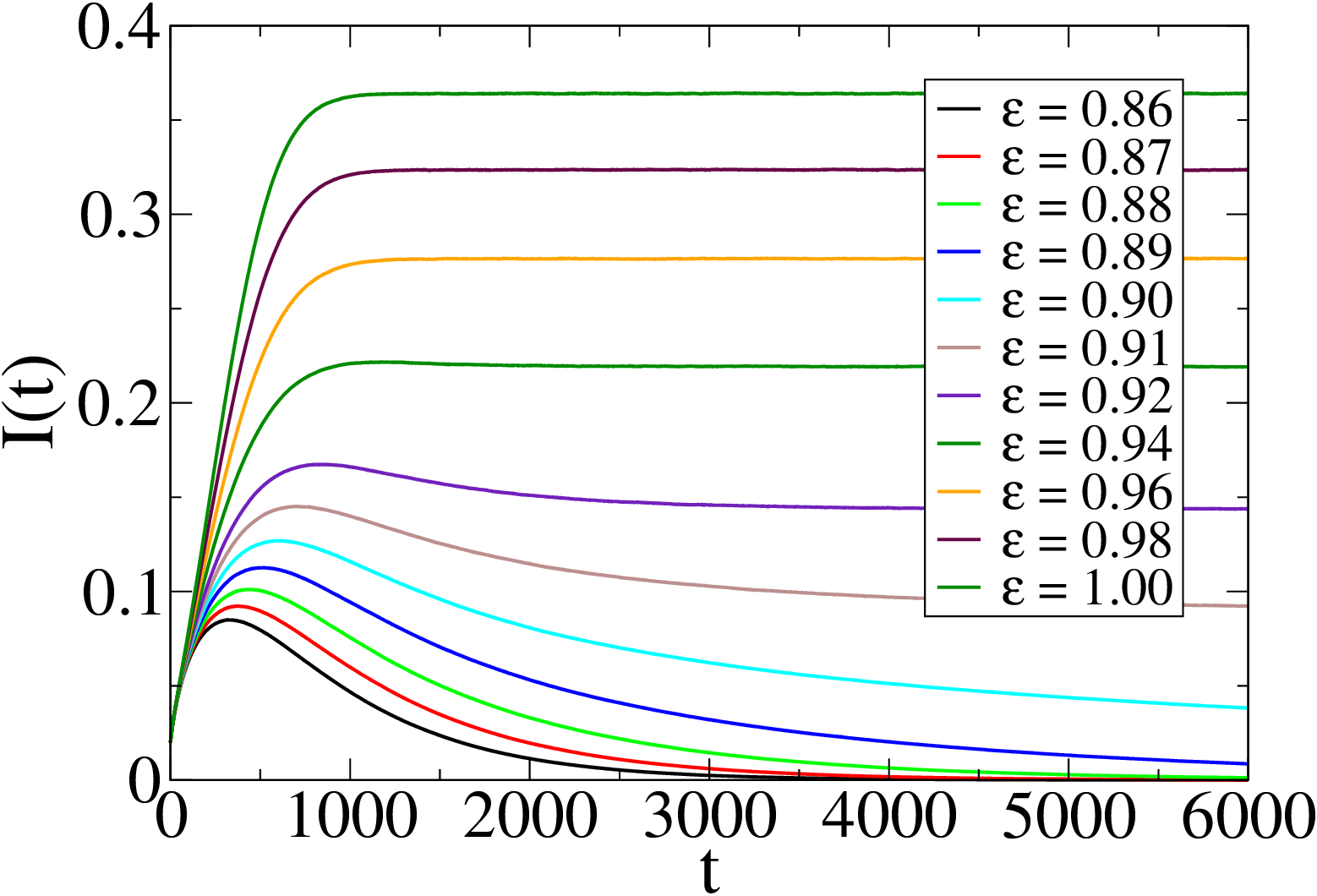}
\hspace{1.0cm}
\includegraphics[width=0.45\textwidth,angle=0]{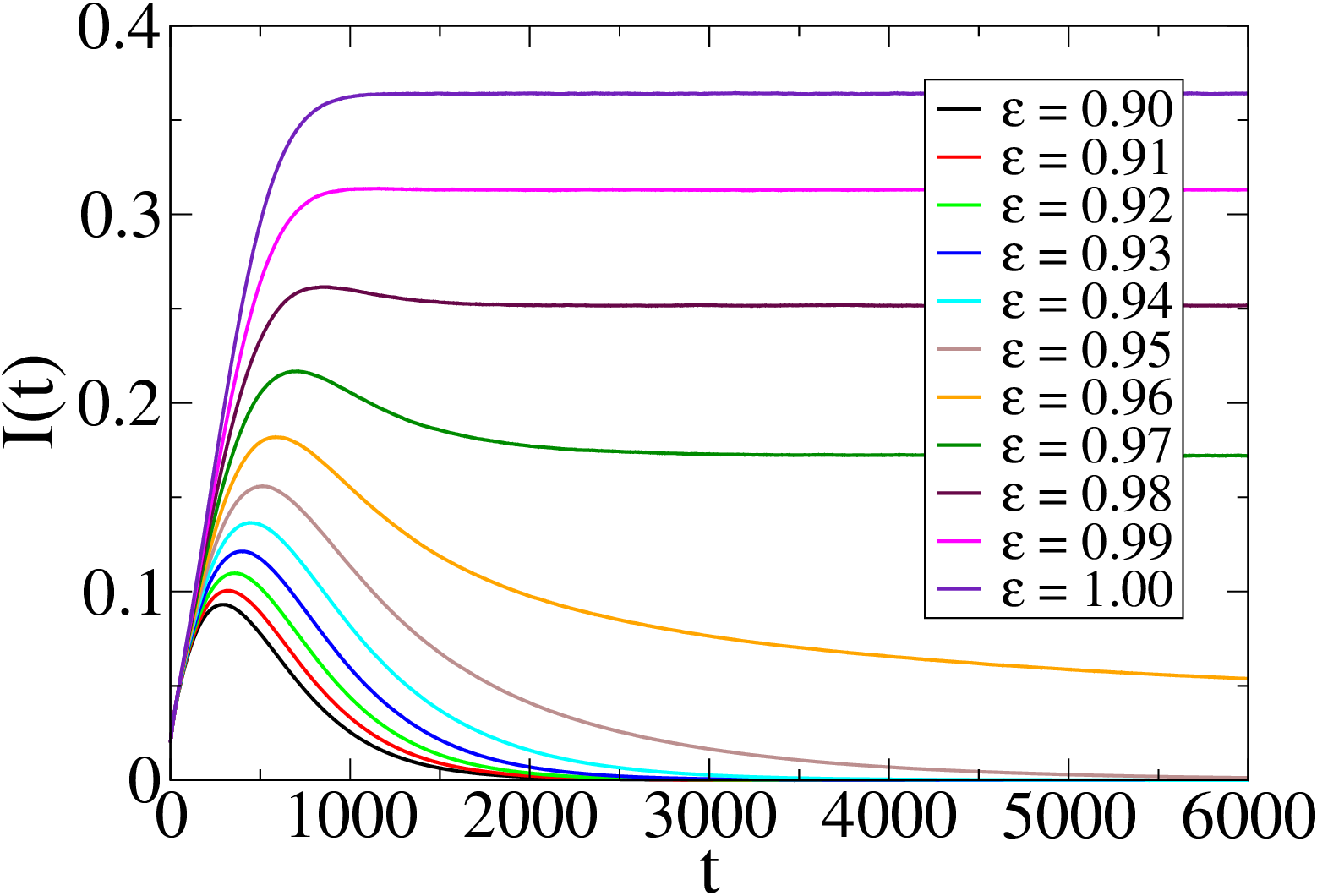}
\end{center}
\protect\caption{Density of Infected individuals $I(t)$ as a function
  of time $t$ for the model defined on a square lattice. The
  parameters are $l=2$ (left side, from top to bottom: $\epsilon=1.00,
  0.98, 0.96, ..., 0.86$) and $l=5$ (right side, from top to bottom:
  $\epsilon=1.00, 0.99, 0.98, ..., 0.90$) and typical values of
  $\epsilon$. Data are averaged over $200$ realizations. We have used
  in these simulations $L=1000$, $\alpha=0.05$ and $\lambda_{0}=0.1$.}
\label{fig7}
\end{figure}

\qquad In this section we will analyze the same model presented in the
last section, but now it will be defined on regular d-dimensional
lattices of linear sizes $L$. The algorithm to simulate the problem is
as follows: (i) at each time step, each Infected individual $j$
returns to the Susceptible state with probability $\alpha$; (ii) at
the same time, each Susceptible individual $j$ becomes Infected with
probability $m\,\lambda(j,t)/z$, where $m=0,1,...,z$ is the number of
infected nearest neighbors of the individual $j$, $z$ is the
coordination number of the d-dimensional lattice ($z=2d$) and
$\lambda(j,t)$ is the infection probability of the individual $j$ at a
certain time step $t$. After each transition $I\to S$, the infection
rate of an individual decreases in the form
$\lambda\to\epsilon\;\lambda$.

Initially, we will consider a square lattice. Thus, we have $N=L^{2}$
individuals, and each individual interact with four neighbors
($z=4$). We have observed that the results do not depend strongly on
the lattice size. Thus, we simulated populations of size up to
$N=10^{6}$ individuals, i.e., for linear sizes up to $L=1000$, with
probabilities $\alpha=0.05$ and $\lambda_{0}=0.1$ and different values
of $\epsilon$ and $l$. We have considered that $2\%$ of the
individuals are initially Infected in the population. All results were
averaged over $200$ realizations.

In Fig. \ref{fig7} we show results for the density of Infected
individuals as a function of time for $l=2$ (left side) and $l=5$
(right side). We can see that, at least qualitatively, the results are
the same as in the mean-field limit. However, the critical values
$\epsilon_{c}(l)$ are different. As examples, we can see in
Fig. \ref{fig7} that we have $\epsilon_{c}(l=2)\sim 0.88$ and
$\epsilon_{c}(l=5)\sim 0.95$, whereas we have for the mean-field case
$\epsilon_{c}(l=2)\sim 0.71$ and $\epsilon_{c}(l=5)\sim 0.87$. In
other words, the difference bewteen the two analyses increases for
increasing values of $l$. This is a consequence of the presence of a
topology (neighborhood) in the model.

\begin{figure}[t]
\begin{center}
\vspace{0.5cm}
\includegraphics[width=0.45\textwidth,angle=0]{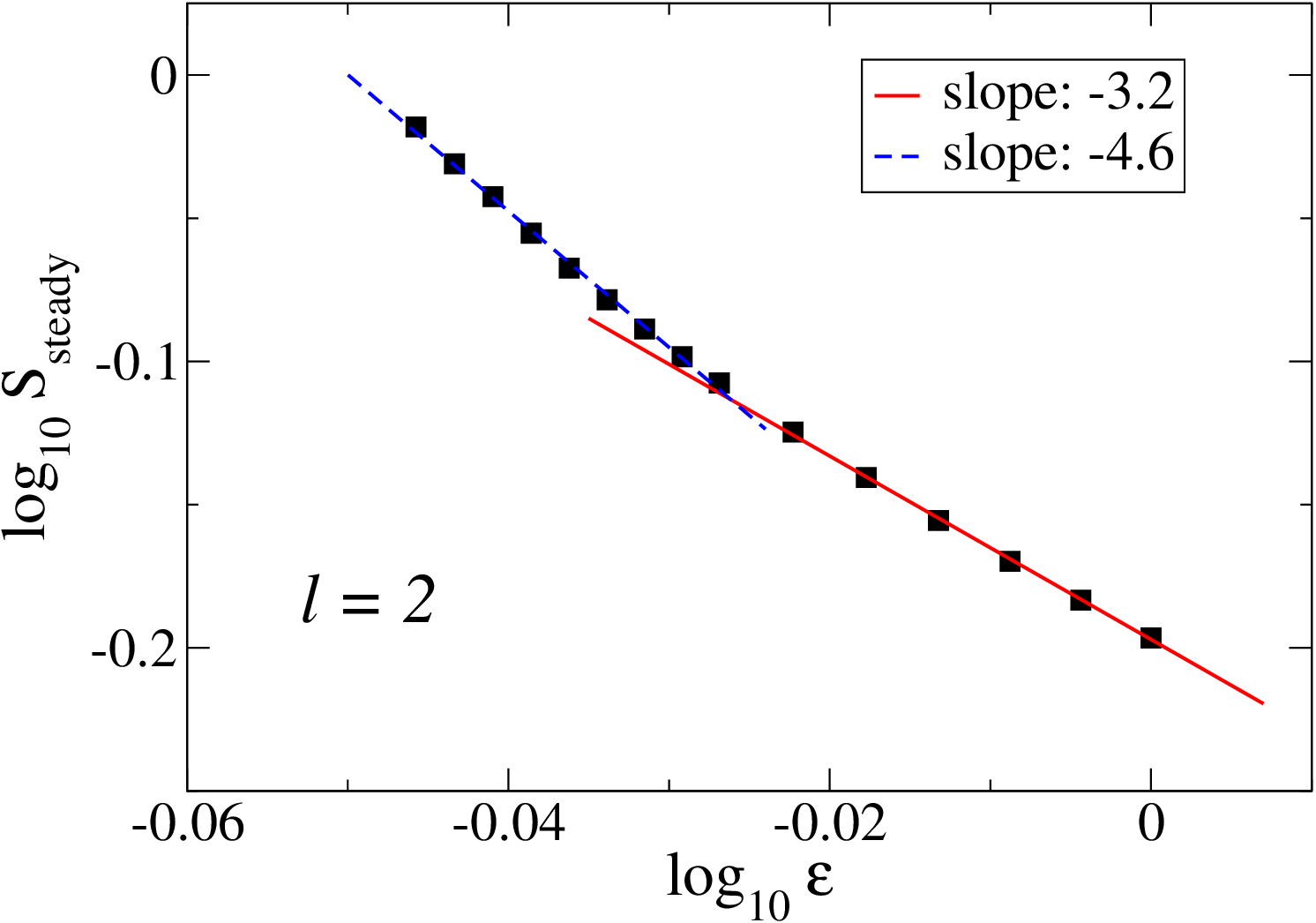}
\hspace{1.0cm}
\includegraphics[width=0.45\textwidth,angle=0]{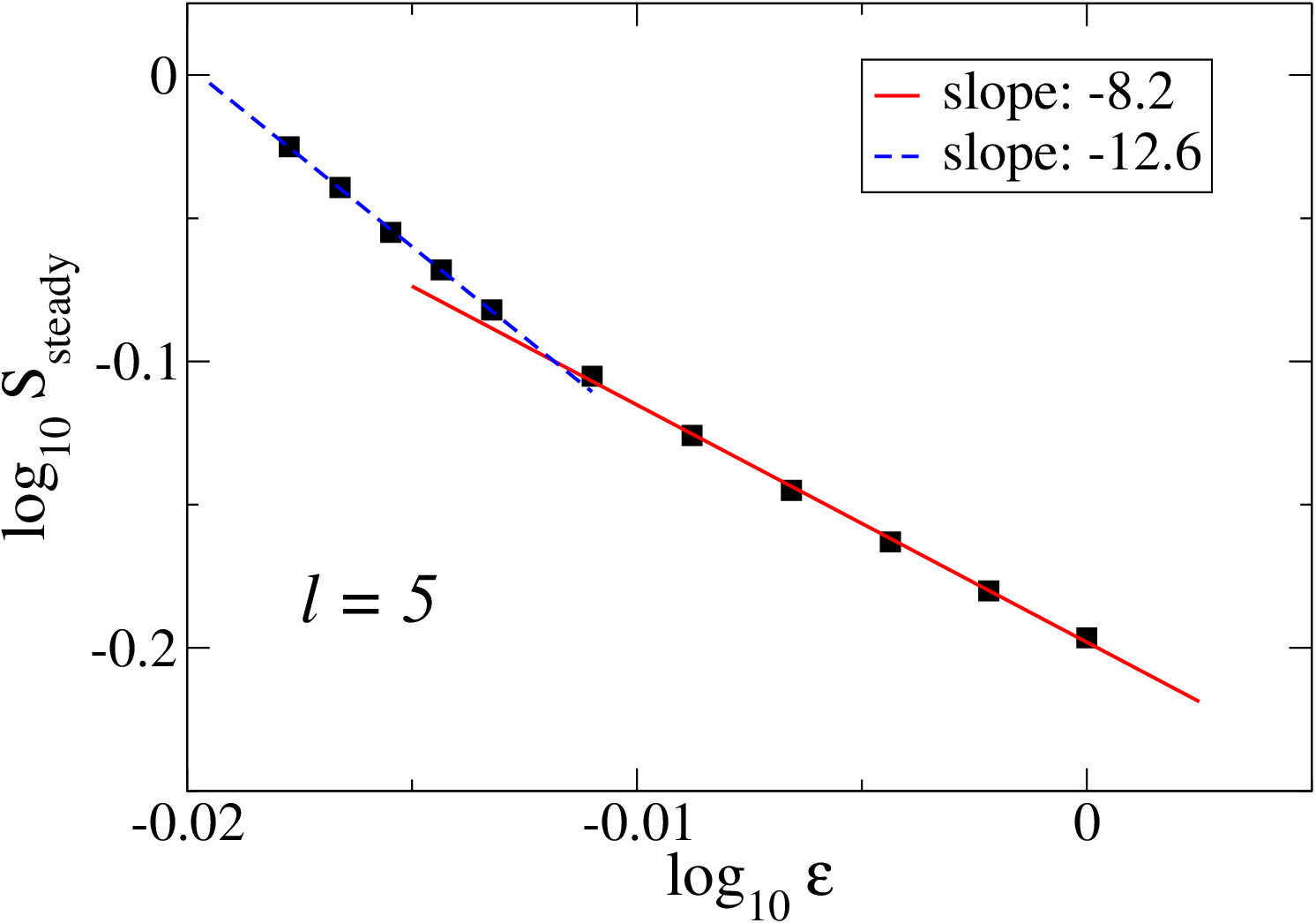}
\end{center}
\protect\caption{Stationary density of Susceptible individuals
  $S_{{\rm steady}}$ as a function of $\epsilon$ in the log-log scale
  for the model defined on a square lattice. The parameters are $l=2$
  (left side) and $l=5$ (right side). Notice that the pure power-law
  behavior of Eq. (\ref{eq6}) is not observed in the 2D case. However,
  there are two distinct power-law behaviors (full and dashed
  lines). We have used in these simulations $L=1000$, $\alpha=0.05$
  and $\lambda_{0}=0.1$. Each point is averaged over 200
  realizations.}
\label{fig8}
\end{figure}

\begin{figure}[t]
\begin{center}
\vspace{0.5cm}
\includegraphics[width=0.45\textwidth,angle=0]{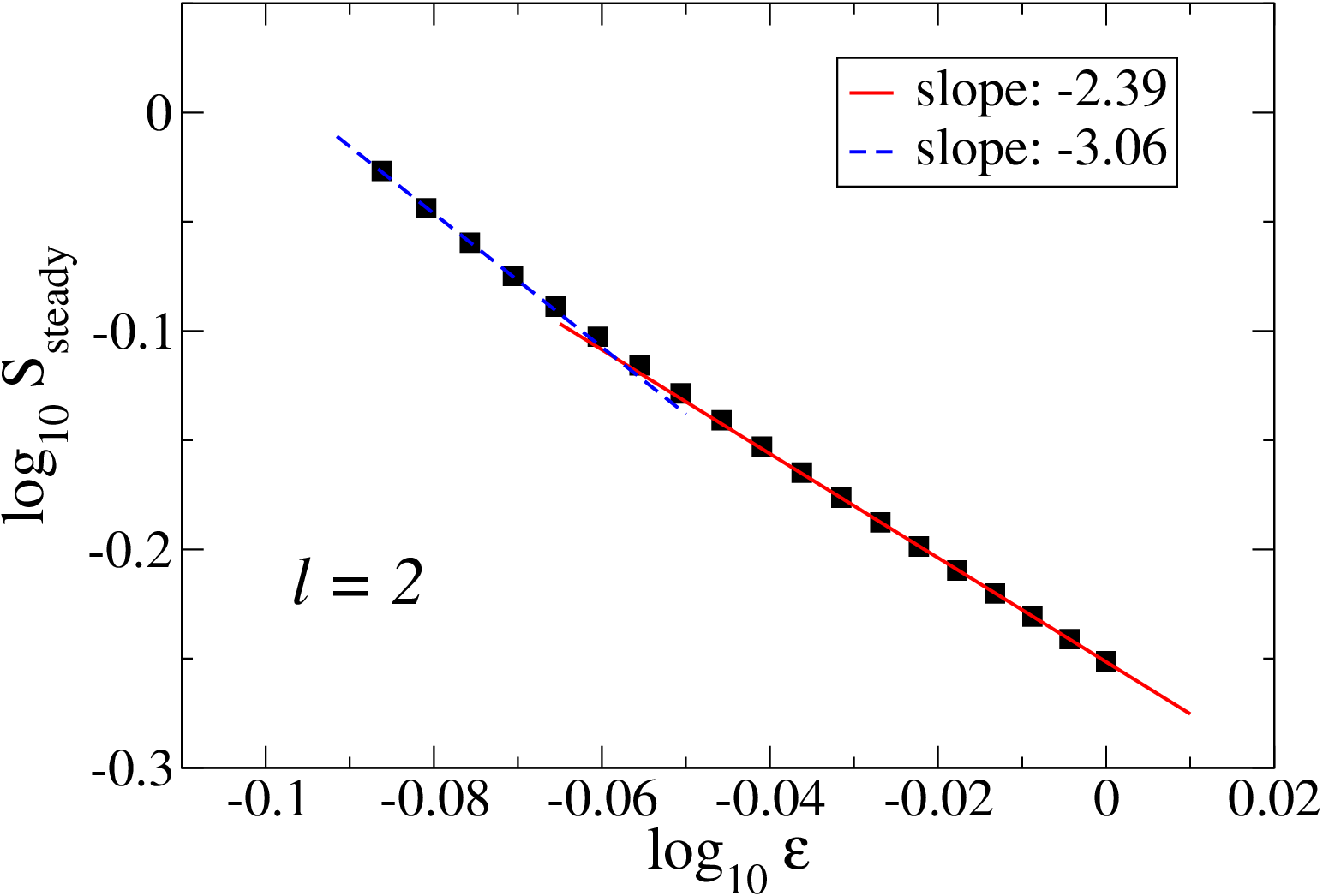}
\hspace{1.0cm}
\includegraphics[width=0.45\textwidth,angle=0]{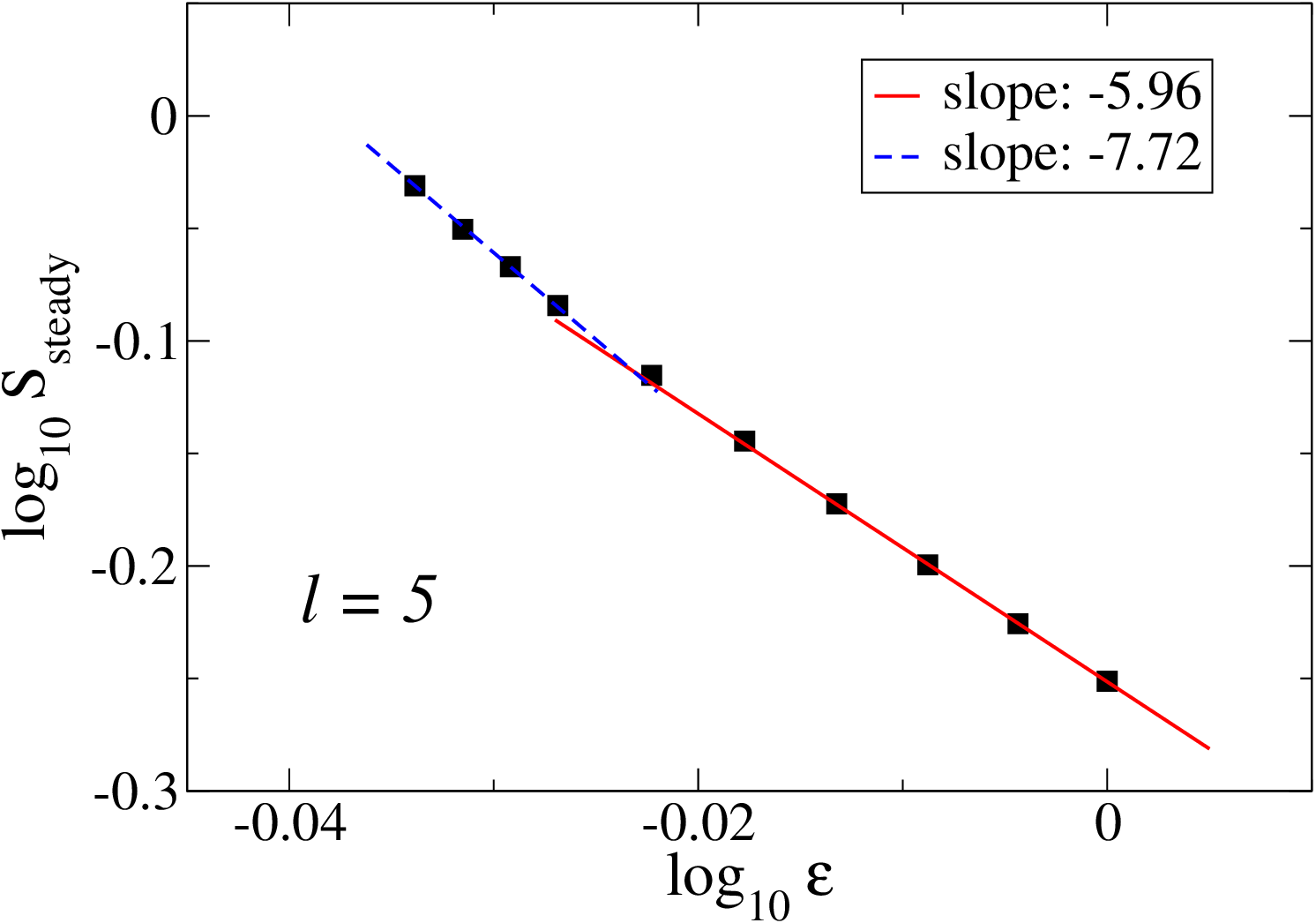}
\end{center}
\protect\caption{Stationary density of Susceptible individuals
  $S_{{\rm steady}}$ as a function of $\epsilon$ in the log-log scale
  for the model defined on a simple cubic lattice. The parameters are
  $l=2$ (left side) and $l=5$ (right side). Notice that the pure
  power-law behavior of Eq. (\ref{eq6}) is not observed in the 3D
  case. However, there are two distinct power-law behaviors (full and
  dashed lines). We have used in these simulations $L=50$,
  $\alpha=0.05$ and $\lambda_{0}=0.1$. Each point is averaged over 200
  realizations.}
\label{fig9}
\end{figure}

\begin{figure}[t]
\begin{center}
\vspace{0.5cm}
\includegraphics[width=0.45\textwidth,angle=0]{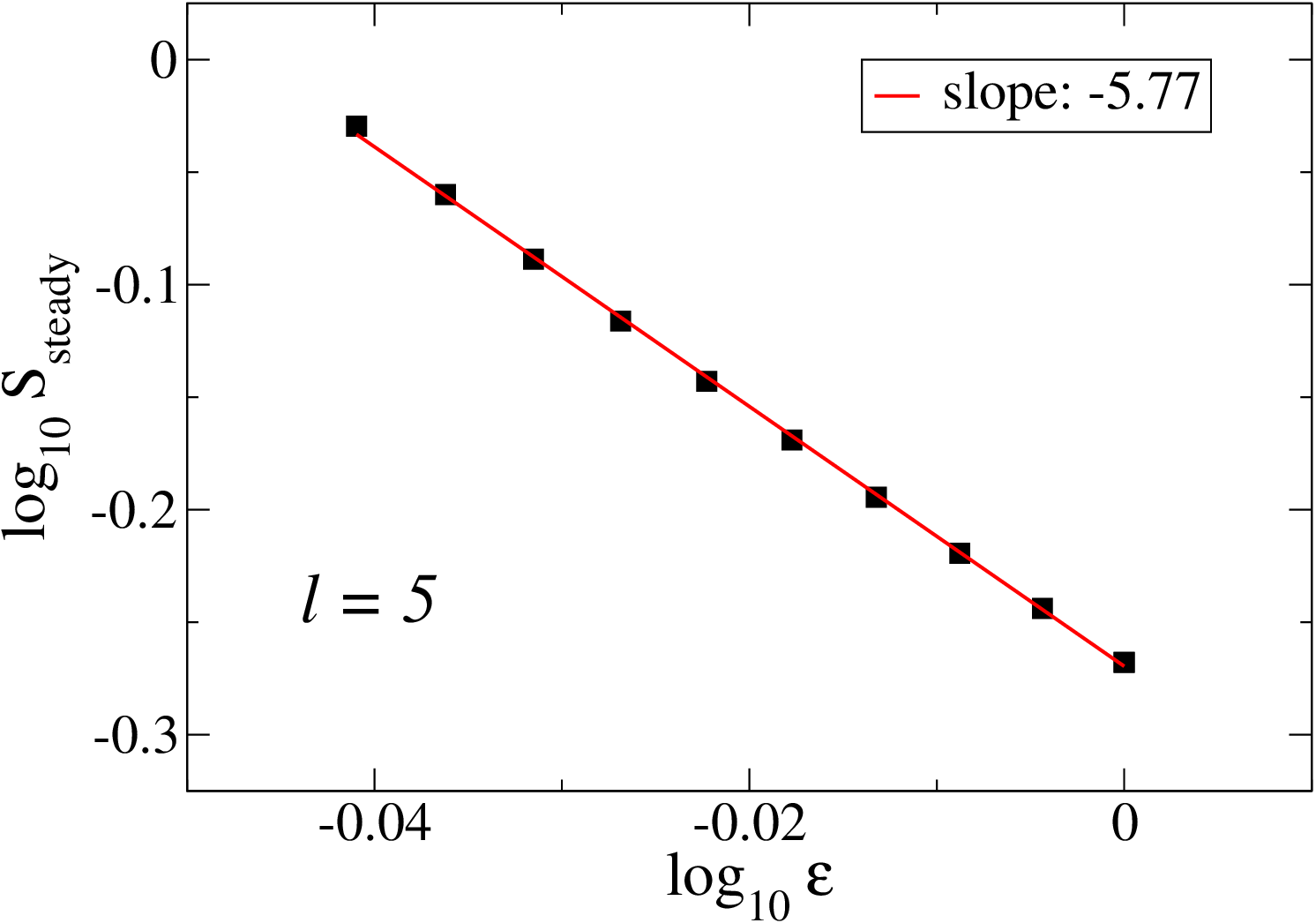}
\hspace{1.0cm}
\includegraphics[width=0.45\textwidth,angle=0]{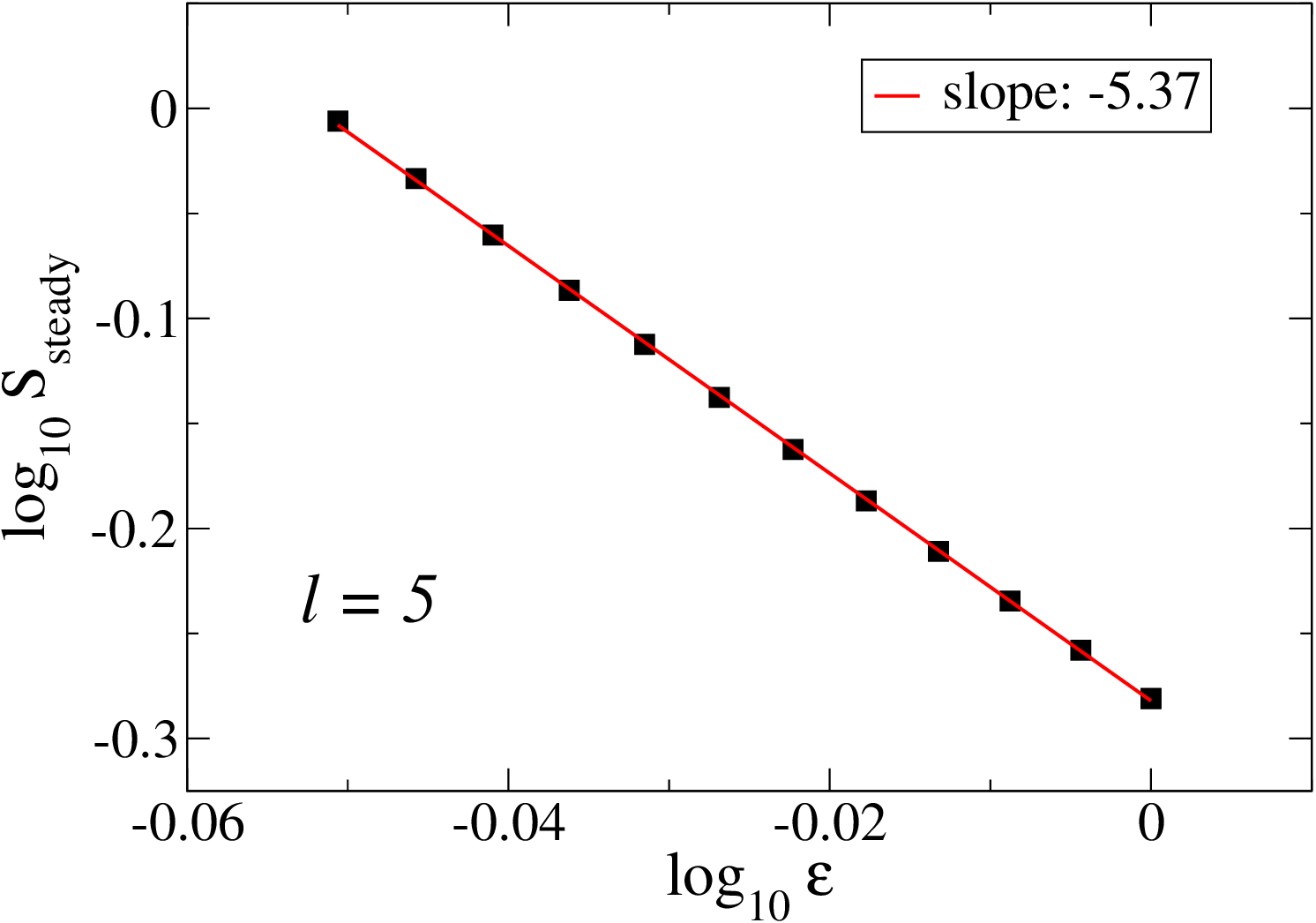}
\end{center}
\protect\caption{Stationary density of Susceptible individuals $S_{{\rm steady}}$ as a function of $\epsilon$ in the log-log scale for the model defined on hypercubic lattices with dimensions $d=4$ (left side) and $d=6$ (right side). Notice that the pure power-law behavior of Eq. (\ref{eq6}) is observed for $d\geq 4$, and that for $d=6$ we have essentially the same behavior as in the mean-field case, i.e., we have $S_{{\rm steady}}\sim \epsilon^{-l}$. The parameters used in the simulations are $\alpha=0.05$, $\lambda_{0}=0.1$, and the lattice sizes considered were $L=20$ (for $d=4$) and $L=8$ (for $d=6$). Each point is averaged over 200 realizations.}
\label{fig10}
\end{figure}

Following the procedure of the previous section, we can analyze the
behavior of the stationary density of Susceptible individuals $S_{{\rm
    steady}}$ as a function of $\epsilon$, for different values of
$l$. We show in Fig. \ref{fig8} results for $l=2$ (left side) and
$l=5$ (right side). We can observe deviations of the single power-law
behavior given by Eq. (\ref{eq6}). The $S_{{\rm steady}}$ values
follow power laws with two different exponents: one for values of
$\epsilon$ near the critical point (bigger slope, dashed lines in
Fig. \ref{fig8}), and another to intermediary and large values of
$\epsilon$, with a smaller slope (full lines in Fig. \ref{fig8}). The
same behavior was observed for the system defined on a simple cubic
lattice ($z=6$), as we can see in Fig. \ref{fig9}. This may be viewed
as a consequence of a small number of neighbors. In fact, if we
consider higher-dimensional lattices we recover a similar mean-field
behavior, i.e., a power-law dependency of the $S_{{\rm steady}}$ on
the parameter $\epsilon$ (see Fig. \ref{fig10}). Considering
for example $l=5$, the numerical results give us $S_{{\rm
    steady}}(l=5)\sim \epsilon^{-5.77}$, $S_{{\rm steady}}(l=5)\sim
\epsilon^{-5.55}$, $S_{{\rm steady}}(l=5)\sim \epsilon^{-5.37}$ and
$S_{{\rm steady}}(l=5)\sim \epsilon^{-5.34}$ for $d=4$, $d=5$, $d=6$
and $d=7$, respectively. In other words, these results suggest that
for $d\geq 6$ the system presents a similar behavior observed in the
mean-field level, with a difference less than $7\%$ to the mean-field
 exponent [see Eq. (\ref{eq6})]. A similar behavior was also observed
for other values of the parameter $l$ as $l=2$ and $l=10$.

\begin{figure}[t]
\begin{center}
\vspace{0.5cm}
\includegraphics[width=0.55\textwidth,angle=0]{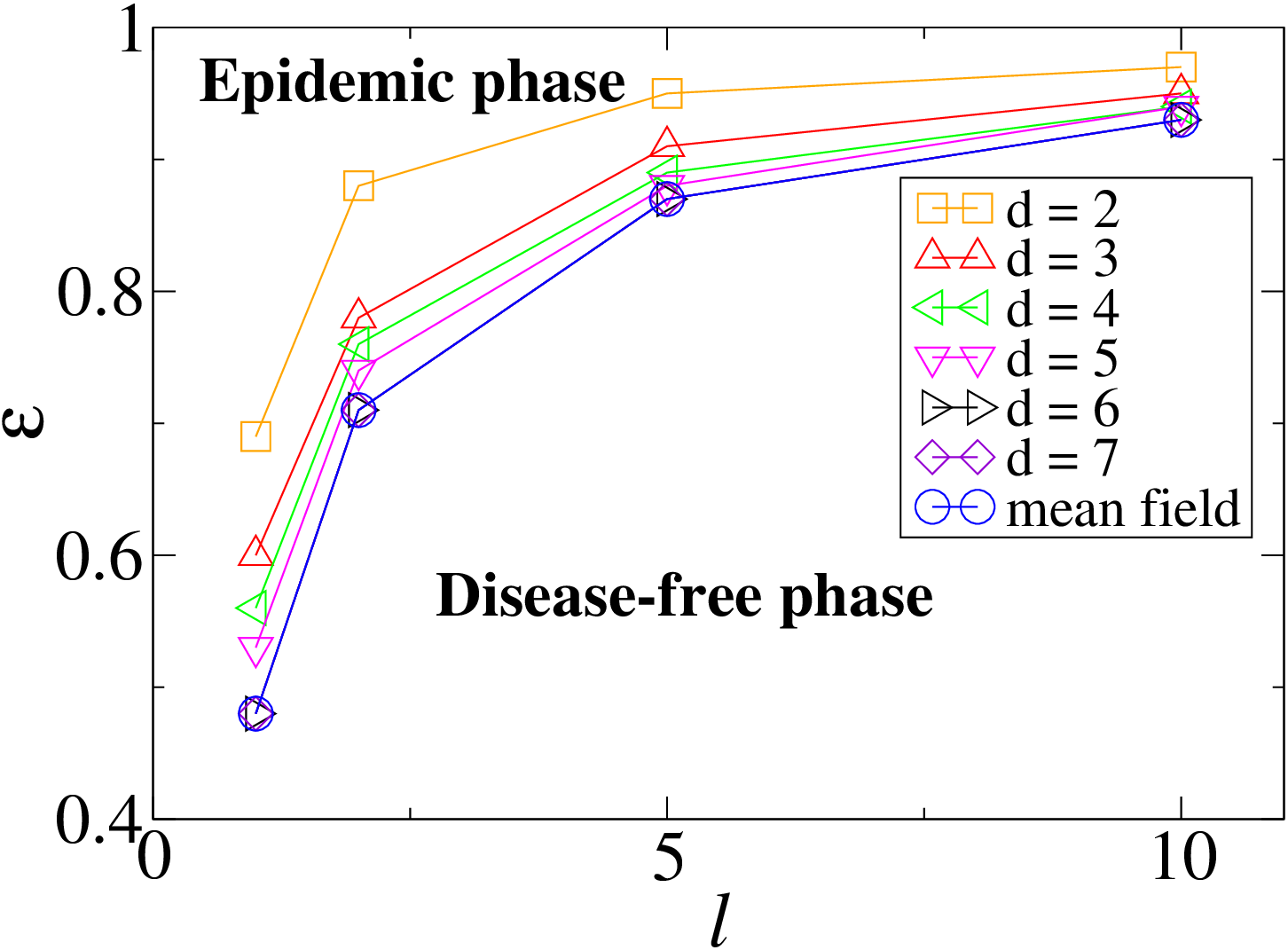}
\end{center}
\protect\caption{Comparative phase diagram of the model in the plane $\epsilon$ versus $l$ for different dimensionalities $d$. The symbols were estimated from the simulations. Notice that the Epidemic phase increases for increasing values of $d$, and that the critical values $\epsilon_{c}(l)$ for $d\geq 6$ and for the mean-field approach are the same. The parameters are $\alpha=0.05$ and $\lambda_{0}=0.1$.}
\label{fig11}
\end{figure}

This picture becomes more clear if we estimate the critical values
$\epsilon_{c}(l)$ for different dimensions $d$. In Fig. \ref{fig11} we
exhibit the phase diagram of the model for some values of $d$. Notice
that the Epidemic phase increases for increasing values of the
dimensionality $d$. This is a consequence of the increasing number of
neighbors (or the coordination number $z$) on higher-dimensional
lattices: it is easier to infect an individual that has more
neighbors. It can be also observed in Fig. \ref{fig11} that the
critical values of $\epsilon_{c}(l)$ for $d=6$, $d=7$ and for the
mean-field case are indistinguishable, which reinforces that the upper
critical dimension of the model is $d=6$, as in the SIR model 
\cite{hinrichsen,newman_SIR}. This is due to the frozen state of the 
individuals with very low infectivity that results from multiple reinfections.

\section{Conclusions}

\quad In this work we studied a modified Susceptible-Infected-Susceptible
(SIS) model in which we have considered that each individual in the
population that recovered from the disease decreases his probability
of reinfection. This decrease occurs a maximum number $l$ of times for
each individual. This dynamics can be relevant to describe diseases
such as flu, for which our chance to be reinfected decreases with age.

Firstly, we have analyzed the problem in the mean-field limit. In this
case, every individual interact with all others, and we studied the
problem with numerical simulations and analytical
calculations. Considering the initial infection rate $\lambda_{0}$
(for the transition $S\to I$) and the recovering rate $\alpha$ (for
the transition $I\to S$), we have found a power-law dependency between
the stationary density of susceptible individuals $S_{{\rm steady}}$
and the parameter $\epsilon$ that controls the decaying of the
infection rate in the form $S_{{\rm
    steady}}=(\alpha/\lambda_{0})\epsilon^{-l}$.  In addition, we
showed that the system undergoes a phase transition at critical values
$\epsilon_{c}(l)=(\alpha/\lambda_{0})^{1/l}$ separating a phase where
the disease reaches a finite fraction of the population (for
$\epsilon>\epsilon_{c}$) from a phase where the disease does not spread
out (for $\epsilon\leq\epsilon_{c}$). All results were confirmed by
Monte Carlo simulations.

Considering $d$-dimensional regular lattices, we have studied the
model only with numerical simulations. The evolution of the density of
Susceptible and Infected individuals is qualitatively similar to the evolution
obtained in the mean-field case, but the presence of a neighborhood modify some
characteristics of the model. In particular, the critical values
$\epsilon_{c}(l)$ are greater than in the mean-field case for $d<6$,
which implies that the epidemic phase decreases for decreasing values
of $d$. However, for $d=6$ and $d=7$ we have obtained the same values
of the critical points $\epsilon_{c}(l)$ as in mean-field
calculations. The stationary density of susceptible individuals
$S_{{\rm steady}}$ depends on the parameter $\epsilon$ in a power-law
form only for $d>3$, whereas for $d=2$ and $d=3$ we have the
combination of two power laws. These numerical results suggest that
the upper critical dimension of the model is $d=6$ as in the SIR
model. This is possibly due to the frozen state of the individuals
with very low infectivity that results from multiple reinfections.

Thus, the mean-field analytical calculations describe qualitatively
well the model, providing us the phenomena that are expected to be
observed, i.e., it predicts the phase transition and the power-law
dependency between some quantities of interest. However, it fails
quantitatively, as it is common in mean-field approximations, because
predicts different values of the critical points and different
power-law exponents.


\section*{Acknowledgments}
The authors acknowledge financial support from the brazilian funding agency CNPq.



\section*{References}
\begin{thebibliography}{40}

\bibitem{perelson_review} A. M. Smith, A. S. Perelson, Wiley
  interdisciplinary Reviews: Systems Biology and Medicine {\bf 3}, 429
  (2011).

\bibitem{perelson_viral_dynamics} P. Baccam, C. Beauchemin, C. A. Macken,
  F. G. Hayden, A. S. Perelson, Journal of Virology 80, 7590
  (2006).

\bibitem{vicsek_experimental} B. G\"onci, V. N\'emeth, E. Balogh, B. Szab\'o,
  A. D\'enes, Z. K\"ornyei, T. Vicsek, PLoS ONE 5(12): e15571 (2010).

\bibitem{anderson} R. M. Anderson, R. M. May, \textit{Infectious
  Diseases of Humans: Dynamics and Control} (Oxford University Press,
  1991).

\bibitem{dickman} J. Marro, R. Dickman, {\it Nonequilibrium Phase
  Transitions in Lattice Models} (Cambridge University Press,
  Cambridge, 1999).

\bibitem{hinrichsen} H. Hinrichsen, Advances in Physics {\bf 49}, 815
  (2000).

\bibitem{newman_SIR} M.E.J. Newman, Phys. Rev. E {\bf 66}, 016128 (2002).

\bibitem{bailey} N. T. J. Bailey, \textit{The Mathematical Theory of
  Infectious Diseases and its Applications} (Hafner Press, New York,
  1975).



\bibitem{derouich} M. Derouich, A. Boutayeb, E. H. Twizell,
  BioMedical Engineering 2:4 (2003).

\bibitem{esteva} L. Esteva, C. Vargas, J. Math. Biol. \textbf{46},
  31 (2003).

\bibitem{nuraini} N. Nuraini, E. Soewono, K. A. Sidarto,
  Bull. Malays. Math. Sci. Soc \textbf{30}, 143 (2007).

\bibitem{rita}
R.M.Z. dos Santos, S. Coutinho, Phys. Rev. Lett. \textbf{87}, 168102 (2001).

\bibitem{baryarama}
F. Baryarama, L. S. Luboobi, J. Y. T. Mugisha, Amer. J. Infect. Diseases \textbf{1}, 55 (2005).

\bibitem{gordo}
I. Gordo, M. G. M. Gomes, D. G. Reis, P. R. A. Campos, PLos ONE 4(3):e4876 (2009).

\bibitem{katriel}
G. Katriel, L. Stone, PLoS Curr. 1: RRN10460 (2009).


\bibitem{cohen} R. Cohen, S. Havlin, D. ben-Avraham, Physical
  Review Letters {\bf 91}, 247901 (2003).

\bibitem{doi}
M. Doi, J. Phys. A \textbf{9}, 1479 (1976).

\bibitem{janssen}
H. K. Janssen, Z. Phys. B \textbf{42}, 151 (1981).

\bibitem{dp_field_theory}
P. Grassberger, A. de la Torre, Ann. Phys. \textbf{122}, 373 (1979).

\bibitem{grassberger}
P. Grassberger, Z. Phys. B \textbf{47}, 365 (1982).

\bibitem{may}
R. M. May, A. L. Lloyd, Phys. Rev. E \textbf{64}, 066112 (2001).

\bibitem{moreno}
Y. Moreno, R. Pastor-Satorras, A. Vespignani, Eur. Phys. J. B \textbf{26}, 521 (2002).

\bibitem{pastor_satorras_pre}
R. Pastor-Satorras, A. Vespignani, Phys. Rev. E \textbf{63}, 066117 (2001).

\bibitem{pastor_satorras_prl}
R. Pastor-Satorras, A. Vespignani, Phys. Rev. Lett. \textbf{86}, 3200 (2001).

\bibitem{havlin_prl}
R. Parshani, S. Carmi, S. Havlin, Phys. Rev. Lett \textbf{104}, 258701 (2010).

\bibitem{boguna}
M. Bogu\~n\'a, R. Pastor-Satorras, A. Vespignani, in \textit{Statistical Mechanics of Complex Networks}, edited by R. Pastor-Satorras, M. Rubi and A. D\'{\i}az-Guilera (Springer Verlag, Berlib, 2003), vol. 625 of \textit{Lecture Notes in Physics}.

\bibitem{sir_plants}
C. A. Gilligan, S. Gubbins, S. A. Simons, Phyl. Trans. R. Soc. Lond. B \textbf{352}, 353 (1997).

\bibitem{perelson_siv}
N. K. Vaidya, R. N. Ribeiro, C. J. Miller, A. S. Perelson, Journal of Virology \textbf{84}, 4302 (2010).





\end{thebibliography}
\end{document}